\def\bphi{\underline{\phi}}
\def\bvphi{\underline{\varphi}}
\def\bchi{\underline {\chi}}
\def\bpi{\underline{\pi}}
\def\bpsi{\underline{\psi}}
\def\bff{\underline{f}}
\def\bJ{\underline{J}}

\documentstyle[12pt,epsf]{article}
\begin{document}
\newcommand{\nd}[1]{/\hspace{-0.5em} #1}
\begin{titlepage}
\begin{flushright}
LA-UR-93-2994 \\
September 1993  \\
\end{flushright}

\begin{centering}
\vspace{.6in}
{\Large {\bf Soliton quantization and internal symmetry}}

\vspace{.5in}
Nicholas Dorey\footnote{Address after
1st October 1993: Physics Department, University College of Swansea,
Swansea, SA2 8PP, UK.} and  Michael P. Mattis \\

\vspace{.1in}
Theoretical Division, T-8 MS B285, Los Alamos National Laboratory, \\
Los Alamos, NM 87545, USA. \\
\vspace{.5in}
James Hughes\footnote{Present Address: Physics Department,
Michigan State University, East Lansing, MI 48823, USA.}\\
\vspace{.1in}
Lawrence Livermore National Laboratory, Livermore, CA 94551, USA. \\
\vspace{.8in}
{\bf Abstract} \\
\vspace{.05in}
\end{centering}
{\small We apply the method of collective coordinate quantization to a
model of solitons  in  two spacetime  dimensions with a  global $U(1)$
symmetry.    In particular we  consider  the  dynamics  of the charged
states associated with  rotational excitations of  the  soliton in the
internal   space and   their interactions   with  the   quanta  of the
background field (mesons). By solving a system of coupled saddle-point
equations   we effectively  sum   all tree-graphs contributing to  the
one-point Green's function  of the meson  field in the background of a
rotating soliton.   We  find that   the resulting one-point   function
evaluated between soliton states  of definite $U(1)$ charge exhibits a
pole on the meson mass shell and we extract the corresponding S-matrix
element for the decay of an excited state via the emission of a single
meson using the standard LSZ  reduction formula. This S-matrix element
has a natural  interpretation in terms  of an effective Lagrangian for
the  charged  soliton states with  an  explicit Yukawa coupling to the
meson field. We calculate the leading-order semi-classical decay width
of  the  excited  soliton states  discuss   the  consequences of these
results for the hadronic decay of the $\Delta$ resonance in the Skyrme
model.}
\end{titlepage}
\section{Introduction}
\paragraph{}
Soliton solutions in  classical  field  theory are parametrized  by  a
finite number of collective coordinates.  In general, each  collective
coordinate corresponds to a continuous symmetry of the theory which is
broken by the  soliton. In addition  to the usual translational modes,
the soliton can  also possess degrees of   freedom corresponding to  a
compact internal  symmetry  of the  theory. In   this case, a  special
feature arises  when the theory   is  quantized and  rotations in  the
internal  space  give  rise to an   infinite tower  of  excited states
transforming  according to successively  higher representations of the
internal  symmetry group   \cite{RW}.  An important   example of  this
phenomenon  occurs for  the   Skyrme  model \cite{skyrme} where    the
isorotational excitations  of  the  skyrmion are identified  with  the
baryon states of large-$N$ QCD, the two lowest states corresponding to
the nucleon  and the $\Delta$  multiplets \cite{ANW,AN}.  In  the full
quantum theory, the soliton states interact with the elementary quanta
of the  background  field which we will  refer  to as  `mesons'. These
mesons carry the conserved  charge associated with the global internal
symmetry  and  transitions between  consecutive soliton  states in the
tower can occur via the emission or absorption of  a single meson.  In
the case of   the Skyrme model,   the quanta of  the chiral  field are
naturally identified with the pions and a simple  example of the above
transitions is    the hadronic   decay  of  the   $\Delta$  resonance;
$\Delta\rightarrow  N+\pi$.  In this paper  we  present  a method  for
calculating  the  leading semi-classical contribution to  the S-matrix
for  these  processes. For clarity  we illustrate  this method  in the
context of the  simplest possible model, a scalar  field theory in two
spacetime  dimensions having     an  unbroken  $U(1)$  symmetry.    In
particular, we calculate the leading-order decay widths of the excited
soliton states in  this model.  The  extension of  our results to  the
$SU(2)$ Skyrme model will be presented elsewhere \cite{wip}.
\paragraph{}
Beyond the  calculation of a specific decay  amplitude, we intend this
paper to be the first in a series clarifying what  we perceive to be a
confused state  of  the  literature on meson-soliton  interactions.  A
useful way to illustrate this confusion is to consider the division of
the  meson field into a  classical  part, corresponding to the soliton
background,   and      a      fluctuating   part     according      to
$\phi=\phi^{cl}+\delta\phi$. This division, which is a generic feature
of soliton quantization, also serves  as a convenient characterization
of  the  different contributions     to the   meson-soliton   S-matrix
considered by  various authors. In  the Skyrme case,  Diakanov, Petrov
and Pobylitsa \cite{DP}  considered the classical contribution to  the
two-point Green's  function of   the  meson  field, which  we   denote
schematically   as $\langle\phi^{cl}\,\phi^{cl}\rangle$. They   argued
that it  contributes a  Born term  to the amplitude  for  pion-nucleon
scattering at  the same   order as   the background  scattering  terms
considered by several   authors \cite{HS,MK}.   The latter terms   are
extracted from the two-point function for the fluctuating field, which
can be  written   as  $\langle \delta\phi\,\delta\phi  \rangle$.   The
argument for a classical contribution to the meson-soliton S-matrix is
quite  general  and dates back   to  the early  literature on  soliton
quantization \cite{GJ,CG}.     In  particular,    the   characteristic
exponentially   decaying     tail    of  the   soliton     background,
$\phi^{cl}(\underline{x})$,    as   $|\underline{x}|\rightarrow\infty$
dictates that  the matrix element of the  background  field exhibits a
pole  in momentum space. The  corresponding residue can be interpreted
as a  point-like coupling of  the mesons to  the  soliton states which
leads   to a Born    contribution  to  the meson-soliton    scattering
amplitudes \cite{GSII}.
\paragraph{}
In principle, the same point-like interaction should mediate the decay
of an excited soliton  state  to a  lower one via   the emission of  a
single   meson.  However, the  classical   contribution  to the matrix
element of the  meson field between initial  and final  states has the
following general form as $|\underline{x}|\rightarrow\infty$:
\begin{eqnarray}
\langle f|\phi^{cl}(\underline{x})|i \rangle
&\sim & {\rm FT}\left[\frac{\delta(k_{0}-
\Delta E)}{|\underline{k}|^{2}+m^{2}}\right]
\label{pole1}
\end{eqnarray}
where $\Delta E$ is the splitting between the states, $m$ is the meson
mass and FT denotes Fourier transform.  In order  to contribute to the
S-matrix element for the  decay, the momentum-space one-point function
should exhibit a pole at the real value of  the spatial momentum given
by    the    mass-shell      condition:    $|\underline{k}|^{2}=\Delta
E^{2}-m^{2}$.   Clearly,   the   one-point   function   determined  by
(\ref{pole1}), exhibits a  pole at an imaginary  value  of the spatial
momentum  given by $|\underline{k}|^{2}=-m^{2}$.    As a  result,  the
classical contribution to  the one-point Green's function cannot, {\em
by itself}, contribute to the physical  decay amplitude.  In contrast,
several attempts to calculate  the amplitude for $\Delta$-decay purely
from the contribution   of the fluctuating  field, $\delta\phi$,  have
also appeared in the Skyrme model literature \cite{verschelde}.  These
authors identify  a linear term in the  effective action for the field
fluctuations around the  rotating skyrmion background. This term gives
a   contribution  to the matrix  element  of   the  pion field between
different baryon states, which  we can write schematically as $\langle
f|\delta\phi|i \rangle$.  As usual, the  field $\delta\phi$ is defined
subject to  functional constraints which render  it  orthogonal to the
soliton zero-modes.  The    fact  that the physical decay    amplitude
obtained in this  way is highly dependent upon  the precise choice  of
this constraint  is  a  strong indication that    the results of  Refs
\cite{verschelde} are incorrect.
\paragraph{}
In fact, the correct interpretation of the linear term in $\delta\phi$
is as a  sign that one  is expanding about  the {\em wrong}  classical
background field $\bphi^{cl}$, namely the time-dependent configuration
obtained by rotating   the  static skyrmion.  More  specifically,  the
linear  term reflects the fact that  this field configuration does not
satisfy the full  time-dependent equation of  motion of the model.  In
this connection it is useful to  note that the analogous configuration
obtained by translating the   soliton does not satisfy   this equation
either.    For the  case of   uniform  translation,  this  is simply a
consequence of the relativistic invariance of the theory; a soliton is
an extended object and hence as it translates with a constant velocity
$v$, the appropriate  solution of the  classical equation of motion is
Lorentz contracted  by   a  factor  $\sqrt{1-v^{2}/c^{2}}$.    In  the
two-dimensional model considered by Gervais,  Jevicki and Sakita (GJS)
\cite{GJS}, similar  linear terms  arise in  the effective action  for
fluctuations of the field around the background of a translating kink.
These terms are precisely due to the mismatch  between the static kink
solution and its Lorentz contracted counterpart.  By  solving a set of
coupled saddle-point equations, GJS successfully recovered the Lorentz
contracted kink as the true  stationary point of the effective action.
The saddle-point contribution    to  the action  corresponds   to  the
infinite sum  of all   vacuum   tree diagrams in   the   weak-coupling
perturbation theory for the fluctuating field.
\paragraph{}
In this paper we solve an analogous  set of saddle-point equations for
the  case  of   an   internally   rotating soliton   in    the  $U(1)$
model. Despite the  fact that there is no  symmetry of the model which
plays   a role   analogous  to  that   of   Lorentz symmetry   in  the
translational case, the  resulting saddle-point configuration, denoted
$\bar{\phi}^{cl}$, has a simple  form. Our main result is conveniently
expressed in  terms   of the  matrix   element of  this new  classical
background      field        between     soliton   states.          As
$|\underline{x}|\rightarrow\infty$ we have,
\begin{eqnarray}
\langle f|\bar{\phi}^{cl}(\underline{x})|i \rangle
& \sim & {\rm FT}\left[\frac{\delta(k_{0}-
\Delta E)}{|\underline{k}|^{2}+m^{2}-\Delta E^{2}}\right]
\label{pole2}
\end{eqnarray}
Comparing (\ref{pole1})   and   (\ref{pole2}), we see  that   the sole
modification of    the classical   one-point   function due     to the
replacement $\phi^{cl}\rightarrow\bar{\phi}^{cl}$ is to shift the pole
in the static soliton background to  exactly the position required for
a  physical contribution to  the  S-matrix. The  new one-point Green's
function (\ref{pole2})  is equal to  the old one (\ref{pole1}) plus an
infinite  number of tree-diagrams  in the  perturbation theory for the
fluctuating field.
\paragraph{}
The first part of the paper is a straightforward generalization of the
methods of GJS to  the case of an  internal degree of  freedom.  After
introducing the $U(1)$ model in  Section 2, in  Section 3 we perform a
canonical  transformation   of  the  path   integral  variables  which
separates out a collective coordinate $\theta(t)$ corresponding to the
$U(1)$ phase angle, from the remaining field degrees  of freedom.  The
expansion of the field around the rotating soliton background contains
linear  terms analogous to  those  found in  the   Skyrme case and  in
Section 4 we solve the  corresponding set of saddle-point equations to
find the  correct  stationary point of the  effective  action. We show
explicitly  that  the resulting field   configuration  is free of  any
ambiguities related to the choice of the functional constraint.  After
replacing   the  meson  field  by  its    saddle-point   value in  the
path-integral  for the   one-point function,  the remaining functional
integral over the collective coordinate is equivalent  to a problem in
Hamiltonian quantum mechanics.   In Section 5   we solve this  problem
exactly, paying careful attention to the problem of operator ordering;
the S-matrix element for the decay of an excited soliton state and the
corresponding decay width is given in  Section 6. We find the S-matrix
element is  formally  equivalent  to that given  at  tree-level  by an
effective Lagrangian  in     which the charged    soliton   states are
represented by  fermionic fields with  an explicit Yukawa  coupling to
the  meson field. Finally, in  Section 7, we discuss some consequences
of these results for the Skyrme model.
\section{The $U(1)$ model}
\paragraph{}
A simple  example  of solitons  having  a compact  internal  degree of
freedom occurs for  the case  of a   real  two-component scalar  field
$\bphi=(\phi_1,\phi_2)$ in two space-time dimensions,
\begin{equation}
{\cal L}=\frac{1}{2}(\partial_{\mu} \bphi)\cdot
(\partial^{\mu} \bphi)-\frac{m^{2}}{2}|\bphi|^{2}- W[\bphi,\sigma]
\label{lag1}
\end{equation}
where the potential $W$ is  chosen so that the Lagrangian (\ref{lag1})
has    an  unbroken  global   $U(1)$  symmetry, $\bphi\rightarrow{\cal
M}(\theta)\bphi$, where
\begin{equation}
{\cal M}(\theta)=
\left(\begin{array}{cc} \cos\theta & \sin\theta \\ -\sin\theta &
\cos\theta \end{array} \right)
\label{rot}
\end{equation}
and    the   resulting   classical  field     equations  admit static,
finite-energy solutions.  In general,  for the theory to satisfy  both
these conditions,  $W$  must also  depend  on at least  one additional
scalar field  $\sigma$     and  its derivatives,   which   couples  to
$\bphi$. Models of  this type were  first considered  by Rajaraman and
Weinberg \cite{RW}. These authors  investigated the specific  case for
which
\begin{equation}
W=\frac{a}{2}|\bphi|^{2}(\sigma^{2}-\mu^{2}/\lambda^{2})
+\frac{h}{4}|\bphi|^{4} - \frac{1}{2}(\partial_\mu \sigma)^{2}-
\frac{\mu^{2}}{2}\sigma^{2}+\frac{\lambda}{4} \sigma^{4}
\label{rw}
\end{equation}
and  demonstrated  the  existence and  energetic  stability of soliton
solutions of the required  form for a certain  range of the Lagrangian
parameters.  Although  the additional field  is necessary to stabilize
the solution, it does not transform under the  $U(1)$ symmetry and its
presence does  not affect the outcome  of our  analysis; hence we will
suppress it in the following discussion.
\paragraph{}
We begin by assuming that the classical field equation,
\begin{equation}
\frac{\partial^{2} \phi_{i}}{\partial t^{2}} -
\frac{\partial^{2} \phi_{i}}{\partial x^{2}} + m^{2}\phi_{i}+
\frac{\delta W}{\delta \phi_{i}}=0
\label{tdep}
\end{equation}
admits a static  soliton  solution, $\bphi^{cl}(x)$.  We will  need to
impose the additional technical requirement (see  Appendix B) that the
soliton     points    in       the    same     internal     direction,
$\bphi^{cl}/|\bphi^{cl}|\equiv\hat{\bphi}_{S}$  at  each    point   in
space.    This condition is  always     true for Lagrangians such   as
(\ref{rw}) where the potential $W$ depends only  on $|\bphi|$. In this
case a global  $U(1)$ transformation  suffices  to rotate the  soliton
into    the    first    component     of    the   field     vector  as
$\underline{\phi}^{cl}=(\phi_{S},0)$, which is the convention we adopt
from  now on. In a theory   with unbroken symmetry,  the finite energy
condition implies that the  soliton profile, $\phi_{S}(x)$, must go to
zero rapidly at spatial infinity. In this respect the static solutions
considered here are  different from the  kink configurations found  in
other  two-dimensional    field  theories  which  interpolate  between
different vacua  at  left and  right spatial  infinity. The asymptotic
behavior  of the profile can be  found by linearizing  the equation of
motion; writing $\phi_{S}=\phi_{S}(x;m)$  to emphasize the  parametric
dependence  of   the  solution   on  the  meson  mass   $m$,  we  find
$\phi_{S}(x;m)\rightarrow   A\exp(-m|x|)$  as  $|x|\rightarrow\infty$.
The coefficient $A$    depends on  any  dimensionless ratios   of  the
Lagrangian  parameters   and must  be  determined  by solving the full
non-linear equation of motion.
\paragraph{}
Starting from the particular solution
$\bphi^{cl}$, the general static
solution is generated by spatial translations and $U(1)$ rotations,
\begin{equation}
\bphi^{cl}(x;X,\theta)={\cal M}(\theta)\bphi^{cl}(x-X)
\label{general}
\end{equation}
In addition to these time-independent solutions, equation (\ref{tdep})
also has solutions which rotate in the internal  space with a constant
angular   velocity  $\omega$. These   are   easily  found  by  writing
$\bphi(x,t)={\cal         M}(\omega    t)\tilde{\bphi}(x)$        with
$\tilde{\bphi}=(\phi_{\omega},0)$.   The resulting    equation     for
$\phi_{\omega}$ is,
\begin{equation}
-\frac{\partial^{2}\phi_{\omega}}{\partial x^{2}} +
\frac{\delta W}{\delta \phi_{\omega}} +
(m^{2}-\omega^{2})\phi_{\omega}=0
\label{roteq}
\end{equation}
For $\omega=0$, equation  (\ref{roteq}) reduces to the static equation
of motion which is solved by the  soliton profile $\phi_{S}(x;m)$. The
effect  of non-zero angular velocity  is simply to shift the effective
mass term in the field equation and the corresponding solution is just
$\phi_{\omega}=\phi_{S}\left(x;\sqrt{m^{2}-\omega^{2}}\right)$.    For
$\omega>m$, the asymptotic behavior of the solution is oscillatory and
the  solution is no longer localized   in space.  In the corresponding
quantum  theory, soliton states  of sufficiently high internal angular
momentum are   unstable and  can decay  into  a lower  state  via  the
emission of a  single meson. As  we will demonstrate below,  the field
configurations  $\phi_{\omega}$  for   suitable   values of  $\omega$,
dominate  the    path-integral  describing    this  process   in   the
semi-classical limit.    In this case,   the  oscillating tail of  the
configuration  is  naturally identified    with the  asymptotic  meson
emitted in the decay.
\paragraph{}
The zero-mode of  the  soliton  associated with infinitesimal   $U(1)$
rotations  is   obtained    by  differentiating  the  static  solution
(\ref{general})  with respect  to the symmetry  parameter $\theta$. In
the chosen  coordinate system, this  rotational zero-mode  is given by
$\bpsi_{R}=(0,\phi_{S})$. The  moment  of inertia of the  soliton with
respect to  internal  rotations can  be expressed   in  terms  of  the
zero-mode as,
\begin{equation}
\Lambda=\int dx\, \bpsi_{R}(x)\cdot \bpsi_{R}(x).
\label{massdef}
\end{equation}
By  analogy with the Skyrme model  we will describe the semi-classical
limit of the  theory in  terms  of a  dimensionless coupling  constant
$1/N$.  The conventional  semi-classical limit is obtained  by setting
$\phi_{S}\sim\sqrt{N}$ in which case  $\Lambda \sim N$.  For the model
of Rajaraman and Weinberg, this  behavior follows from taking both the
dimensionless  ratios    $h/m^{2}$    and   $\lambda/\mu^{2}$   to  be
$O(1/N)$. Expanding the field  around the static soliton background in
the $N\rightarrow\infty$ limit as $\bphi(x,t)=\bphi^{cl}(x)+
\bchi(x,t)$ we find,
\begin{equation}
{\cal L}[\bphi]={\cal L}[\bphi^{cl}]+ \frac{1}{2}\dot{\bchi}\cdot
\dot{\bchi}+\frac{1}{2}\bchi^{T}\hat{L}\bchi + V(\bchi)
\label{expand}
\end{equation}
where the operator of quadratic fluctuations is
\begin{equation}
\hat L_{ij}
=\left(-\frac{\partial^2}{\partial x^2}+m^2\right)\delta_{ij}
+{1\over2}{\delta^2W\over\delta\phi_i\delta\phi_j}[\underline{\phi}^{cl}]
\label{fluc}
\end{equation}
\paragraph{}
In the limit $|x|\rightarrow\infty$, the potential contribution to the
fluctuation         operator  goes      to        zero   rapidly,  and
$\hat{L}_{ij}\rightarrow  (-\partial_{x}^{2}+m^{2})\delta_{ij}$  which
is   just  the  free inverse   propagator  for field  quanta   of mass
$m$. However,  as  expected, the   full operator  ${\hat  L}$ has  two
vanishing  eigenvalues and so cannot  be inverted  directly. Using the
$U(1)$ invariance of  the potential $W$, it  can easily be  shown that
one of   the zero  eigenfunctions  of  $\hat{L}$  coincides   with the
rotational  zero-mode    $\bpsi_{R}$ as  expected.   The   other  zero
eigenfunction is associated with infinitesimal translations and can be
obtained in  a  similar  way by  differentiation  with respect  to the
soliton center of mass coordinate, $X$.
\section{Collective coordinate quantization}
\paragraph{}
The  zero-modes discussed in the   previous  section lead directly  to
infra-red  divergences in the  naive perturbation expansion around the
soliton  background  coming  from the  Lagrangian  (\ref{expand}).  In
order to  construct a finite  perturbation theory  it is  necessary to
remove  these  zero-modes from  the  spectrum by  introducing explicit
collective coordinate fields $X(t)$ and $\theta(t)$.  In this paper we
will adopt the path-integral  quantization method of Gervais,  Jevicki
and Sakita (GJS) in which the collective coordinates are introduced by
performing a canonical  transformation  of the dynamical  variables in
the   phase-space    path integral \cite{GJS,GS}.      This method was
originally developed   in the context  of a  kink configuration in two
dimensions with a   single  translational degree of freedom   and this
section is  a straightforward  generalization  of their method  to the
case of  an internal collective coordinate.   In the following we will
focus exclusively  on the dynamics of   internal soliton rotations and
will henceforth ignore the  translational  motion of the soliton.  The
collective coordinate quantization   of solitons with  a translational
zero mode has been extensively  discussed in the literature and offers
no new  features in  the present case.   Our  analysis  can easily  be
extended  to include  the translational degree   of freedom using  the
methods described  in  \cite{GJS}.  In  particular,  the inclusion  of
translations leads   to recoil corrections   to the  S-matrix elements
calculated below which are suppressed by  factors of $1/N$ relative to
the leading-order results.
\paragraph{}
The transition  amplitude  between initial  and  final soliton states,
described by wavefunctions $\Psi_{i}$ and  $\Psi_{f}$, at times $t=-T$
and $t=+T$ can be written as a phase-space path integral,
\begin{equation}
T_{fi}[\bJ]=\int {\cal D}\bphi{\cal D}\bpi
\Psi_{f}^{*}[\bphi(x,+T)]\Psi_{i}[\bphi(x,-T)]
\exp\left(i\int d^{2}x\, \bpi\cdot \dot{\bphi}-{\cal H}(\bpi,\bphi)
+ \bJ\cdot\bphi \right)
\label{tran}
\end{equation}
with the Hamiltonian density,
\begin{equation}
{\cal H}=\frac{1}{2}\bpi \cdot\bpi + \frac{1}{2}\bphi'\cdot\bphi'
+\frac{m^{2}}{2}|\bphi|^{2}+W(\bphi)
\end{equation}
where the  conjugate field momenta, $\bpi=(\pi_{1},\pi_{2})$ are given
by   $\pi_{i}={\delta{\cal L}}/{\delta  \dot{\phi}_{i}}$  and  we have
included  an external   source $\bJ$  which  couples  to  $\bphi$.  In
Section 5 below we will   identify soliton states of definite   $U(1)$
charge and specify the wavefunctions in (\ref{tran}) accordingly.  The
Green's functions describing  the meson emission and absorption  which
accompanies  transitions between these states  can then be obtained by
functional differentiation with respect to the source.
\paragraph{}
Following the standard method of collective coordinate quantization we
introduce $\delta$-function    constraints      orthogonalizing    the
fluctuations of the  field   around  the  soliton  background  to  the
rotational  zero  mode by inserting   the  following identity into the
integrand of the transition amplitude,
\begin{equation}
\int {\cal D}\theta(t) {\cal D}P(t) J_{\theta}J_{P}
\delta(F_{1}[\theta;{\bf
\bphi}])\delta(F_{2}[\theta,P; {\bf \bphi}, {\bf \bpi}])=1
\label{delta}
\end{equation}
where  $J_{\theta}=\delta F_{1}/\delta\theta$ and $J_{P}=\delta  F_{2}
/\delta P$  and  the   angular collective  coordinate   $\theta(t)$ is
restricted  to lie in the range  $[0,2\pi]$. In  the following, $P(t)$
will be  interpreted as  the internal  angular  momentum conjugate  to
$\theta(t)$ and we will later find that it is restricted to a discrete
set   of values    which label   the   physical soliton   states.  The
constraints $F_{1}$ and $F_{2}$ are chosen as
\begin{equation}
F_{1}(t)=\int dx\, \underline{f}(x)\cdot [{\cal M}^{T}(\theta(t))\bphi(x,t)-
\bphi^{cl}(x))]
\label{f1}
\end{equation}
\begin{equation}
F_{2}(t)=\Lambda^{-1} \int dx\,
\underline{f}(x)\cdot [{\cal M}^{T}(\theta(t))\bpi(x,t)-
\bpi^{cl}(x))]
\label{f2}
\end{equation}
where  $\bpi^{cl}$ will  be  defined  below.   The introduction of   a
$\delta$-function constraint in  this context is somewhat analogous to
the Faddeev-Popov gauge-fixing procedure in  Yang-Mills theory and the
precise  choice   of the   function $\bff(x)$   which   appears in the
constraint is frequently referred to as the choice  of a `gauge'.  The
analogy is a good one to the extent that  all physical quantities such
as S-matrix elements should be `gauge-invariant' at  each order in the
semi-classical  expansion.  In  general,   any choice  for  the  gauge
function which  has  non-zero functional overlap  with  the rotational
zero-mode is  sufficient   to   remove   the corresponding   infra-red
divergences  occurring in  the  naive perturbation  theory around  the
soliton background.   The  simplest choice   is $\bff=\bpsi_{R}$ which
corresponds to the  so-called `rigid  gauge'.  We  will not choose   a
specific  gauge  for this calculation as  we   seek to demonstrate the
gauge-invariance of the resulting  S-matrix elements.  For  notational
convenience we will restrict  our  attention to  gauges for  which the
vector $\bff$ is parallel to $\bpsi_{R}$ at each point in space and we
also   fix the  overall normalization   of these functions   to be the
same. We therefore set $\bff=(0,f(x))$ and impose
\begin{eqnarray}
\int dx\, f(x)\phi_{S}(x)\neq 0 & \qquad{} \qquad{} &
\int dx\, [f(x)]^{2} = \Lambda
\label{conds}
\end{eqnarray}
However, it can easily be checked that our conclusions are not altered
if these restrictions are relaxed
\paragraph{}
We now  make a  non-linear  change of variables  in the  path-integral
introducing  the      rotated        field        $\tilde{\bphi}={\cal
M}^{T}(\theta(t))\bphi$.   This change of  variables can be thought of
as a transformation from the  laboratory frame to the body-fixed frame
of the  rotating soliton.  In general,  given any quantity $A$  in the
laboratory frame,   we will denote  the  corresponding quantity in the
body-fixed frame as $\tilde{A}$.  In   terms of the new variables  the
Jacobian factor $J_{\theta}$ is given  by\footnote{In the following we
define  a  {\em scalar} cross  product  of a   pair of two-dimensional
vectors by $\underline{a}\times\underline{b}=
\epsilon^{ij}a_{i}b_{j}$.},
\begin{equation}
J_{\theta}=\int dx\, \underline{f} \times \tilde{\bphi}
\label{jac}
\end{equation}
The  corresponding     transformation  of  the   field   momentum  is,
$\bpi(x,t)={\cal  M}(\theta(t))(\bpi^{cl}+\tilde{\bpi})$,  where   the
`classical' piece of the  momentum, $\bpi^{cl}$, which also appears in
the constraint (\ref{f2}), must be chosen carefully to ensure that the
above  change of variables  constitutes  a  canonical  transformation.
More  precisely, we demand  that the Legendre  term in the phase-space
path integral retain its canonical form in terms of the new variables,
\begin{equation}
\int dx\, \bpi\cdot\dot{\bphi}=P\dot{\theta}+\int dx\, \tilde{\bpi}
\cdot\dot{\tilde{\bphi}}
\label{legendre}
\end{equation}
and also that the  two Jacobian factors  associated with the change of
variables     cancel  in      the      integrand   of   (\ref{delta}):
$J_{\theta}J_{P}=1$. In Appendix A we show that these two requirements
specify a unique choice  for $\bpi_{cl}$. In  terms of the  body-fixed
frame fields we have,
\begin{equation}
\bpi^{cl}=\frac{1}{J_{\theta}}\left(P-\int dx\,
\tilde{\bpi}\times\tilde{\bphi}\right)\bff(x)
\label{cmom}
\end{equation}
\paragraph{}
The resulting   transition amplitude  can   then  be expressed  as  an
integral  over all  paths in  the   collective coordinate phase  space
$(P(t),\theta(t))$,
\begin{equation}
T_{fi}[\bJ]=\int {\cal D}\theta(t){\cal D}P(t)\Psi^{*}_{f}[\theta(+T)]\Psi_{i}
[\theta(-T)]
\exp\left(i\int_{-T}^{+T} dt P\dot{\theta}\right)
\exp iS[\theta,P;\bJ]
\label{colint}
\end{equation}
where we  have  anticipated the fact  that  the  wave-functions of the
soliton states of fixed $U(1)$  charge are functions of the collective
coordinate only  and do not depend  on the  remaining field degrees of
freedom. The effective action for  a given path, $S[\theta,P;\bJ]$, is
given in  terms  of a  constrained path-integral  over  the body-fixed
frame fields,
\begin{eqnarray}
\exp{iS[\theta,P;\bJ]}&=&
\int{\cal D}\tilde{\bphi}(x,t){\cal D}\tilde{\bpi}(x,t)
\delta\left(\int dx\, \bff\cdot\tilde{\bphi}\right)
\delta\left(\int dx\, \bff\cdot\tilde{\bpi}\right) \nonumber \\
&& \qquad{} \qquad{} \qquad{} \qquad{} \times
\exp\left(i\int d^{2}x\, \tilde{\bpi}\cdot\dot{\tilde{\bphi}}-
\tilde{{\cal H}}[\tilde{\bpi},\tilde{\bphi}]+\bJ\cdot\bphi\right)
\label{effective}
\end{eqnarray}
where the Hamiltonian density for the body-fixed frame fields is found to be,
\begin{equation}
\tilde{\cal H}=\frac{1}{2}\tilde{\bpi}\cdot\tilde{\bpi} +
\frac{1}{2}\tilde{\bphi}'\cdot\tilde{\bphi}' +
\frac{m^{2}}{2}|\tilde{\bphi}|^{2}
+W(\tilde{\bphi})+
\frac{\Lambda}{2J_{\theta}^{2}}\left(P-\int dx\,
\tilde{\bpi}\times\tilde{\bphi}\right)^{2} + \Delta V
\label{restframeH}
\end{equation}
As  with any non-linear   point canonical transformation  of  the path
integral, additional  terms in the   effective potential arise  beyond
those  obtained by naive substitution  in the action.  We have denoted
these terms as $\Delta V$ in (\ref{restframeH}).  In the corresponding
canonical quantization of the  system \cite{Tomboulis}, the additional
terms  are associated with operator ordering  ambiguity presented by a
Hamiltonian  such  as (\ref{restframeH})   which involves  arbitrarily
complicated  products of fields   and their conjugate momenta.  In the
functional  integral  approach adopted here, the  equivalent ambiguity
arises  in   choosing the  appropriate   discretized definition of the
path-integral \cite{midpoint}.  The additional terms in  the potential
are independent of the collective coordinate and its conjugate angular
momentum and contribute to the  renormalization of the soliton profile
function, $\phi_{S}$, at the two-loop level. These terms do not affect
our results for the  S-matrix at leading  order and we  will disregard
them in the following.
\section{Perturbation theory and the saddle-point equations}
\paragraph{}
The first  step in  obtaining  the leading-order  contribution to  the
transition amplitude (\ref{colint})    is to evaluate  the  inner path
integral  over the body-fixed frame   fields (\ref{effective}) in  the
semi-classical limit.    In this   section   we will   accomplish this
directly by  applying the powerful  saddle-point method used by GJS in
the  translational case. Before  applying this  method we will briefly
discuss its relation to  the more familiar weak-coupling  perturbation
theory which was also introduced by GJS in  Ref \cite{GJS}.  Expanding
the  body-fixed frame  field  around the static  soliton background as
$\tilde{\bphi}=\bphi^{cl}+\bchi$, the transition amplitude becomes,
\begin{eqnarray}
T_{fi}[\bJ]&=&\int {\cal D}\theta(t){\cal D}P(t)\Psi^{*}_{f}\Psi_{i}
\int{\cal D}\bchi(x,t){\cal D}\tilde{\bpi}(x,t)
\delta\left(\int dx\, \bff\cdot\bchi\right)
\delta\left(\int dx\, \bff\cdot\tilde{\bpi}\right) \nonumber \\
 & & \qquad{} \qquad{} \qquad{}\qquad{} \qquad{} \qquad{} \qquad{} \qquad{}
\times\exp\left( i\int dt L_{0} + L_{int}+\int dx\, \bJ\cdot\bphi\right)
\label{big}
\end{eqnarray}
where,
\begin{eqnarray}
L_{0} &=& P\dot{\theta}+\int dx\, \tilde{\bpi}\cdot\dot{\bchi}-H_{0}
\label{l0} \\
H_{0} &=& M+ \int dx\,
\left[ \frac{1}{2}\tilde{\bpi}\cdot\tilde{\bpi}
+\frac{1}{2}\bchi^{T}\hat{L}\bchi \right]
\label{h0} \\
L_{int}& = &\frac{\Lambda}{2J_{\theta}^{2}}\left(P-\int dx\,
\tilde{\bpi}\times\tilde{\bphi}\right)^{2}+\int dx\,
V(\bchi)
\label{int}
\end{eqnarray}
\paragraph{}
The analysis given in  the previous Section,  which leads to the above
form  for the  transition  amplitude  is  completely analogous to  the
quantization  of   the  kink system  given    by  GJS. In  particular,
(\ref{big}) is  the natural generalization of corresponding expression
for   the  kink  transition   amplitude given  as   Eqn  (2.14) in Ref
\cite{GJS} and  the   resulting perturbation  theory  has  a   similar
structure.   In both cases, the   $\delta$-function constraints in the
path-integral   effectively  eliminate  the  zero-eigenvalue  of   the
quadratic fluctuation operator  and the propagator for the fluctuating
field can  be obtained by  inverting  this operator on the  functional
subspace  orthogonal to   the    zero-modes. The  expansion   of   the
interaction Lagrangian  in powers of $1/N$  gives rise to  an infinite
series  of vertices for the fluctuating  field some of which depend on
the collective  coordinates.  In   general, these Feynman    rules are
cumbersome and highly   gauge-dependent although, as we  have  already
stressed,  this dependence must eventually    cancel in the  resulting
S-matrix  elements.  In the translational    case,  GJS give  explicit
expressions  for  the propagators and  the first  few vertices  in the
`rigid gauge' where the constraint function is chosen to coincide with
the translational zero mode.  The lowest-order vertex which depends on
the kink momentum is a one-point  vertex which corresponds to a linear
term in the effective action for  the fluctuating field (see Figure 3a
of Ref  \cite{GJS}).  Expanding   the  interaction Lagrangian in   the
present case we find,
\begin{equation}
L_{int}=\frac{\Lambda P^{2}}{2\int dx\, \bff\times\bphi^{cl}}
\left[1 -2\frac{\int dx\, \bff\times\bchi}{\int dx\, \bff\times\bphi^{cl}} +
O(|\bchi|^{2})\right]
\label{tadpole}
\end{equation}
The second  term in the brackets  is  linear in the  fluctuating field
$\bchi$ and yields an analogous  one-point vertex.  When combined with
the propagators and  the   higher-order vertices,  this   vertex leads
directly to  an   infinite number  of connected   tree-diagrams  which
contribute to the transition amplitude.
\paragraph{}
As discussed in  Section  1, the presence of  a  linear term indicates
that   that the  background field configuration   around  which we are
expanding is not a stationary point of  the effective action.  For the
case  of a uniformly  translating   kink, GJS solved the  saddle-point
equations  for the effective   action and demonstrated  that the  true
stationary point is given by the appropriately Lorentz contracted kink
configuration which  also  solves the  full  time-dependent  classical
field equation  of the  model.    As  usual, the saddle-point   method
corresponds to the reorganization of the perturbation theory as a loop
expansion and,  in this context,  the Lorentz contracted kink is equal
to  the the original   kink configuration  plus  the  sum of all  tree
diagrams with one external leg.  Similarly,  in the case of a rotating
soliton considered here, the simple  static background $\bphi^{cl}$ is
not a stationary point  of the exponent of  (\ref{big}) for $P\neq 0$.
By analogy with the  translational case we might  expect that the true
stationary  configuration would be  given by a suitable time-dependent
solution  of  the original  classical  field equation  for the  $U(1)$
model, Eqn (\ref{tdep}). In  the  following we will demonstrate   that
this is indeed the case and give an explicit expression for the sum of
all tree-diagrams contributing to the  one-point function of the meson
field.
\paragraph{}
In order to apply  the saddle-point method  of  Ref \cite{GJS} to  the
constrained path integral (\ref{effective}),  it is first necessary to
exponentiate the constraints by introducing Lagrange multiplier fields
$\bar{\lambda}(t)$ and $\bar{\nu}(t)$.  In  the absence of  the source
$\bJ$, we have,
\begin{equation}
\exp{iS[\theta,P;0]}=
\int{\cal D}\tilde{\bphi}(x,t){\cal D}\tilde{\bpi}(x,t)
{\cal D}\bar{\lambda}(t){\cal D}\bar{\nu}(t)
\exp\left( i\int d^{2}x\,  \tilde{\bpi}\cdot\dot{\tilde{\bphi}}-
\tilde{{\cal H}}+\bar{\lambda}\bff\cdot\tilde{\bphi}+\bar{\nu}\bff\cdot
\tilde{\bpi}\right)
\label{effectiveII}
\end{equation}
The  saddle-point field   configuration which   provides the  dominant
contribution to $S[\theta,P;0]$ in the  semi-classical limit will be a
stationary point of the effective Lagrangian  for the body-fixed frame
fields   which  appears in    the exponent     of (\ref{effectiveII}).
Returning  to  the  full   expression  for  the  transition  amplitude
(\ref{colint})   and  noticing that  $S[\theta,P;0]$   depends only on
$P(t)$ and  not on  $\theta(t)$, we  see that  the angular momentum is
necessarily a  constant  of  the   collective coordinate   motion  for
$\bJ=0$.  For this reason, we will  restrict our attention to the case
$P(t)\equiv  P$  for which the  corresponding   field configuration is
time-independent. The  static saddle-point equation which follows from
varying   the   exponent of  (\ref{effectiveII})     with  respect  to
$\tilde{\phi}_{i}$ is,
\begin{eqnarray}
\frac{\partial^{2}\tilde{\phi}_{i}}{\partial x^{2}} -m^{2}\tilde{\phi}_{i}
-\frac{\delta W}{\delta
 \tilde{\phi}_{i}}&+&\frac{\Lambda}{J_{\theta}^{2}}\varepsilon^{ij}\tilde
{\pi}_{j}\left(P-\int dx\,
\tilde{\bpi}\times\tilde{\bphi}
\right) \nonumber \\
&&\qquad{}\qquad{}\qquad{}-\frac{\Lambda}{J_{\theta}^{3}}\varepsilon^{ij}
f_{j}\left(P-\int dx\,
\tilde{\bpi}\times\tilde{\bphi}
\right)^{2}+\bar{\lambda}f_{i}=0
\label{phieq}
\end{eqnarray}
where we have used
$\delta J_{\theta}/\delta \tilde{\phi}_{i}=-\varepsilon^{ij}f_{j}$ which
follows from (\ref{jac}). The
corresponding equation for the conjugate field momentum
$\tilde{\pi}_{i}$ is given by
\begin{equation}
\tilde{\pi}_{i}+\frac{\Lambda}{J_{\theta}^{2}}\varepsilon^{ij}\tilde
{\phi}_{j}\left(P-\int dx\,
\tilde{\bpi}\times\tilde{\bphi}
\right) -\bar{\nu}f_{i} =0
\label{pieq}
\end{equation}
The saddle-point  field configurations are  found by solving equations
(\ref{phieq}) and  (\ref{pieq})  simultaneously with   the  constraint
equations,
\begin{eqnarray}
\int dx\, \bff\cdot\tilde{\bphi}=0 & \qquad{} \qquad{} &
\int dx\, \bff\cdot\tilde{\bpi}=0
\label{coneq}
\end{eqnarray}
which follow from varying the Lagrangian with
respect to $\bar{\lambda}$ and $\bar{\nu}$.
\paragraph{}
The solution of  the above system of equations  can  be obtained by  a
series of manipulations  which are completely  analogous to those used
by GJS in the translational case; details are given in Appendix B. The
main result  of   this  analysis  is that  both  the  momentum  field,
$\tilde{\bpi}$,  and  the  Lagrange  multipliers, $\bar{\lambda}$  and
$\bar{\nu}$,   can be  eliminated  leaving  the single gauge-invariant
equation,
\begin{equation}
-\frac{\partial^{2} \tilde{\phi}_{S}}{\partial x^{2}} +m^{2}\tilde{\phi}_{S}
+\frac{\delta W}{\delta \tilde{\phi}_{S}}-
\frac{P^{2}\tilde{\phi}_{S}}
{\left( \int dx\, \tilde{\phi}_{S}^{2} \right)^{2}}=0
\label{mastereq}
\end{equation}
where  $\tilde{\bphi}=(\tilde{\phi}_{S},0)$.   Identifying the angular
velocity   $\omega$  in terms  of   the  collective coordinate angular
momentum  by $\omega=P/\tilde{\Lambda}$   where    the field-dependent
moment of inertia $\tilde{\Lambda}$ is given by,
\begin{equation}
\tilde{\Lambda}=\int dx\, \tilde{\phi}_{S}^{2}
\label{moi}
\end{equation}
we see that  equation (\ref{mastereq}) coincides exactly with equation
(\ref{roteq})   obeyed by  $\phi_{\omega}$,   the  uniformly  rotating
solution of   the    original  time-dependent    equation  of   motion
(\ref{tdep}).   It   follows  that   the  corresponding  solution  for
$\tilde{\phi}_{S}$ can be expressed  in terms of the unrotated soliton
profile $\phi_{S}$ as,
\begin{equation}
\tilde{\phi}_{S}(x;m)=\phi_{S}
\left(x;\sqrt{m^{2}-P^{2}/\tilde{\Lambda}^{2}}\right)
\label{shift}
\end{equation}
where the effective moment of inertia,
$\tilde{\Lambda}=\tilde{\Lambda}(P)$ satisfies the
self-consistency condition,
\begin{equation}
\tilde{\Lambda}=\int dx\, \left[\phi_{S}\left(x;\sqrt{m^{2}-P^{2}/
\tilde{\Lambda}^{2}}\right)\right]^{2}
\label{consistent}
\end{equation}
For  any given potential $W$,   solving this self-consistency equation
requires the full  numerical solution  for  the soliton profile for  a
range of values of $m$.  Fortunately, with some mild assumptions about
the general  behavior  of the soliton  solutions as  the meson mass is
varied, we  are able to  demonstrate the  existence  of a solution for
sufficiently  small  internal  angular momentum.    In  particular, we
assume only that the static soliton moment of  inertia is a continuous
function of the meson mass   with a non-zero limit, $\Lambda_{0}$,  as
the meson   mass is taken   to zero.  We   note that the corresponding
assumptions are certainly  true in the  case of the Skyrme model where
static equation of motion has been solved numerically both in the case
of massless pions \cite{ANW} and in the massive case \cite{AN}. In the
Appendix we give a graphical proof that (\ref{consistent}) necessarily
has a real solution for $P/m<\Lambda_{0}$ as long as these assumptions
are  satisfied.   Expanding  both   sides  of  (\ref{consistent})  and
recalling      that       $\phi_{S}\sim\sqrt{N}$     we   see     that
$\tilde{\Lambda}(P)=\Lambda (1 +   O(1/N))$. In the  following we will
also  assume that this  series gives the correct analytic continuation
of $\tilde{\Lambda}$ to the regime $P/m>\Lambda_{0}$.
\paragraph{}
The  resulting    $U(1)$  family  of  saddle-point  configurations  is
generated     by    rotations    of      the  particular      solution
$\tilde{\bphi}^{cl}=(\tilde{\phi}_{S},0)$.  It  is easy  to check that
these configurations together with the corresponding solutions for the
field momentum,  $\tilde{\bpi}^{cl}$,   and the  Lagrange multipliers,
$\bar{\lambda}^{cl}$  and   $\bar{\nu}^{cl}$,   provide  a  consistent
solution    to the     full     system  of    saddle-point   equations
(\ref{phieq}-\ref{coneq}).     The    energy   of   this  solution  is
conveniently expressed as,
\begin{equation}
E(P)=\int dx\, \tilde{{\cal H}}[\tilde{\bphi}^{cl},\tilde{\bpi}^{cl}]
 =\frac{P^{2}}{2\tilde{\Lambda}}+
\int dx\, {\cal H}[\tilde{\bphi}^{cl},\tilde{\bpi}^{cl}]
\label{mass}
\end{equation}
By expanding
(\ref{mass}) in powers of $1/N$, we obtain,
\begin{equation}
E(P)=M+\frac{P^{2}}{2\Lambda}+O\left(\frac{1}{N^{2}}\right)
\label{expansion}
\end{equation}
Thus, as expected, at leading  order the energy of  the soliton due to
its rotation  is exactly that of a  rigid body with  moment of inertia
$\Lambda$.       The     corresponding     action     is     given  by
$S[\theta,P;0]=2E(P)T$  and represents   the   infinite sum   of   all
connected  vacuum  tree-diagrams  in   the weak-coupling  perturbation
theory outlined  above. We note that  it is  manifestly independent of
the gauge function $\bff$.   In the analysis of  Section 3, the source
term was  introduced only as convenient  device  to obtain the Green's
functions of laboratory frame field.  The saddle-point contribution to
these Green's functions  is given by replacing the  full field by  the
same  configuration which  dominates  the  the source-free  transition
amplitude.  For  this  reason  we  disregard  any  modification of the
saddle-point field caused by the reintroduction of source and write,
\begin{equation}
S[\theta,P;\bJ]=\int_{-T}^{+T}dt\left(-E(P)+\int dx\,
\bJ\cdot\bvphi\right)
\label{action}
\end{equation}
where $\bvphi(x,t)={\cal M}(\theta(t))\tilde{\bphi}^{cl}(x)$ is the
corresponding saddle-point configuration of the laboratory field $\bphi$.
\paragraph{}
The  higher-order  corrections to   this result  can  be obtained   by
expanding  the     body-fixed  frame field around     the saddle-point
configuration                                                       as
$\tilde{\bphi}=\tilde{\bphi}^{cl}+\delta\tilde{\bphi}$.    Because  we
are now expanding  about the true stationary point,  there  will be no
linear term  for  the fluctuating field  $\delta\tilde{\bphi}$  in the
effective action and therefore   the corresponding Feynman  rules will
not contain a one-point vertex. It follows that the corrections to the
one-point function in this  perturbation theory necessarily involve at
least one loop.  As expected, the  new perturbation theory is simply a
reorganization of the old perturbation theory as a loop expansion.  In
this  case, the    loop  expansion is    an improvement  on    the old
perturbation theory in two ways. First,  we have demonstrated that the
zeroth-order contribution in   the loop expansion  is manifestly gauge
invariant. Second, as we  discussed in Section  1, single terms in the
old perturbation theory  do  not contribute to the  S-matrix directly.
In the following we  will  demonstrate that the zeroth-order   Green's
functions  obtained by  replacing the  full  field by its saddle-point
value have  physical poles  in momentum  space  which contribute  Born
terms  to  the meson-soliton S-matrix.  However,   we should point out
that  even the loop   expansion by itself does  not  always provide  a
systematic expansion for the S-matrix elements in powers of $1/N$.  In
the  case of  the two-point function,   an  exact cancellation of  the
leading order occurs and the first non-vanishing contribution contains
not only  the subleading part  of the  Born  terms but also background
scattering terms  which come  from  the  next  order in the  expansion
around the saddle-point \cite{DP}.  Fortunately the one-point function
which we will calculate below is free  from any such cancellations and
unambiguously gives the leading order contribution to the S-matrix for
soliton decay.
\section{The transition amplitude and Green's functions}
\paragraph{}
Applying the static saddle-point analysis of the previous section, the
leading  semi-classical  contribution  to  the  source-free transition
amplitude given by (\ref{colint}) can be written as,
\begin{equation}
T_{fi}[0]=\int {\cal D}\theta(t){\cal D}P(t)\Psi^{*}_{f}[\theta(+T)]\Psi_{i}
[\theta(-T)]
\exp\left(i\int_{-T}^{+T} dt P\dot{\theta}-E(P)\right)
\label{colintII}
\end{equation}
This phase-space path   integral  given above describes  the   quantum
mechanics of  a single degree of freedom  governed by  the Hamiltonian
$H=E(P)$.  As  indicated in  the  previous section,   this Hamiltonian
coincides with that  of a non-relativistic  rigid top at leading order
in $1/N$. Using the standard equivalence between the path integral and
canonical quantization,  we introduce a collective coordinate operator
$\hat{\theta}$    and   its    conjugate   momentum    $\hat{P}$  with
$[\hat{\theta},\hat{P}]=i$.  The  momentum operator $\hat{P}$ commutes
with  the corresponding operator   Hamiltonian  $\hat{H}$ and  so  the
stationary states of the system are labelled by a conserved charge $p$
with $\hat{P}|p\rangle=p|p\rangle$.  Because the collective coordinate
is an   angular variable,  the   charge $p$  is  constrained to  be an
integer; the orthonormalized wavefunctions are
\begin{equation}
\psi_{p}(\theta)=\frac{1}{(2\pi)^{\frac{1}{2}}}\exp(ip\theta)
\label{wvfns}
\end{equation}
The corresponding energy levels
are given by $\hat{H}|p\rangle=E(p)|p\rangle$.
\paragraph{}
The stationary states of $\hat{H}$ correspond to excited states of the
soliton of  definite $U(1)$ charge  and are directly analogous  to the
baryon states  of the Skyrme model.   The transition amplitude between
the  states $|p_{i}\rangle$  and   $|p_{f}\rangle$ can be obtained  by
setting   $\Psi_{i}=\psi_{p_{i}}$  and $\Psi_{f}=\psi_{p_{f}}$  in the
integrand of (\ref{colint}). In the absence of  the source $\bJ$ which
couples the collective   coordinates  to the  background field,  these
states are  necessarily stable and  the resulting transition amplitude
is diagonal.  Choosing an appropriate   overall normalization for  the
path integral (\ref{colintII}) we find,
\begin{equation}
T_{fi}[0]=\delta_{p_{f},p_{i}}\exp [-2iTE(p_{i})]
\label{diagonal}
\end{equation}
However,  the field $\bphi$  interpolates  mesons of  unit charge and,
once its coupling   to the quantum mechanical  degrees  of freedom  is
restored, we expect that transitions  between adjacent soliton  states
will  be  mediated by  the absorption   and  emission of these quanta.
These processes are described  the Green's functions obtained from the
corresponding  transition amplitude   in  the presence  of  the source
$\bJ$.  Using  (\ref{action}),  the saddle-point  contribution to  the
transition amplitude is given by,
\begin{equation}
T_{fi}[\bJ]=\int {\cal D}\theta(t){\cal D}P(t)\psi^{*}_{p_{f}}
[\theta(+T)]\psi_{p_{i}}
[\theta(-T)]
\exp\left(i\int_{-T}^{+T} dt P\dot{\theta}-E(P)+\int dx\,\bJ\cdot\bvphi\right)
\label{colintIII}
\end{equation}
It is  convenient    to   write  the  laboratory-frame    saddle-point
configuration        $\bvphi$      in      terms     of     components
$\varphi_{\pm}=\varphi_{1}\mp  i\varphi_{2}$ corresponding   to  field
quanta of definite   $U(1)$ charge.  Using  the  explicit form  of the
saddle-point solution we have;
\begin{equation}
\varphi_{\pm}(x;P(t),\theta(t))= \exp(\pm i\theta(t))
\phi_{S}\left(x;\sqrt{m^{2}-\omega\left(P(t)\right)^{2}}\right)
\label{comps}
\end{equation}
where   $\omega(P)=P/\tilde{\Lambda}(P)$   is   the effective  angular
velocity.   Arbitrary  Green's functions  for  the fields $\phi_{\pm}$
which interpolate asymptotic quanta  of charge $p=\pm 1$  are obtained
by   differentiating the transition  amplitude  (\ref{colintIII}) with
respect to  the  source components  $J_{\pm}=(J_{1}\mp iJ_{2})/2$. The
general $n$-point function obtained in this way is written as,
\begin{equation}
\langle p_{i},-T|\prod_{a=1}^{n}\phi_{s_{a}}(x_{a},t_{a})|p_{f}, +T
\rangle=\left.\frac{\delta^{n} T_{fi}[\bJ]}{\delta J_{s_{n}}(x_{n},t_{n})\ldots
\delta J_{s_{1}}(x_{1},t_{1})} \right| _{\bJ\equiv 0}
\label{grfs}
\end{equation}
where $s_{a}={\rm   sign}(r_{a})$ with   $r_{a}=\pm  1$   specifies  a
particular  choice  of positive and   negative field components. Using
(\ref{colintIII}), the above   Green's function can  be related   to a
corresponding $n$-point function for the  meson field between  initial
and  final soliton states of   definite orientations, $\theta_{i}$ and
$\theta_{f}$,
\begin{equation}
\langle p_{f},+T|\prod_{a=1}^{n}\phi_{s_{a}}(x_{a},t_{a})|p_{i}, -T
\rangle=\int d\theta_{i}d\theta_{f}\psi^{*}_{p_{f}}(\theta_{f})
\psi_{p_{i}}(\theta_{i})\langle \theta_{f},+T|
\prod_{a=1}^{n}\phi_{s_{a}}(x_{a},t_{a})|\theta_{i}, -T \rangle
\label{gg}
\end{equation}
which is defined as a path integral with fixed boundary conditions at
$t=\pm T$;
\begin{equation}
\langle \theta_{f},+T|\prod_{a=1}^{n}\phi_{s_{a}}(x_{a},t_{a})|\theta_{i}, -T
\rangle=\int{\cal D}P(t)\int_{\theta(-T)=\theta_{i}}
^{\theta(+T)=\theta_{f}} {\cal D}\theta(t)
\prod_{a=1}^{n}\varphi_{s_{a}}(x_{a},t_{a})
\exp\left[i\int_{-T}^{+T}dt P\dot{\theta}
-H\right]
\label{gf}
\end{equation}
We see from  the above expression  that  the momentum $P(t)$  is still
conserved except at  the $n$ points  $t_{a}$ and we have  assumed that
the static saddle-point configuration  of the previous section is also
applicable, to leading order, at these isolated points.
\paragraph{}
In the canonical language, the  path integral (\ref{gf}) is equivalent
to a product  of expectation values of  the operators corresponding to
the  saddle     point    fields;    $\hat{\varphi}_{\pm}     =\exp(\pm
i\hat{\theta})\phi_{S}(x;\sqrt{m^{2}-\omega(\hat{P})^{2}})$.     These
composite operators involve products of  arbitrary powers of $\hat{P}$
and   $\hat{\theta}$ and the  correct   ordering convention for  these
non-commuting factors needs    to  be specified. In contrast    to the
ordering problem which  arises   from  the  non-linear form   of   the
canonical   transformation  discussed   in  Section  3,  the resulting
correction   terms in this    case  explicitly involve the  collective
coordinates and must  be included in order  to correctly determine the
positions  of  the  momentum-space poles   in  the tree-level  Green's
functions.   In  the  path-integral  approach, the  operator  ordering
problem is   replaced by an  equivalent ambiguity,  namely the need to
specify  an appropriate  discretized   definition of   the  functional
integral \cite{schul}.  The  appropriate resolution  of this ambiguity
in   the  context of  soliton  quantization  was given  by Gervais and
Jevicki   \cite{midpoint}.  Following  their analysis   we  choose the
symmetric midpoint formula  for  the discretized path  integral.   The
time interval  $[-T,T]$    is   divided   into $N+1$    equal   steps,
$t_{(k)}=-T+k\epsilon$   for  $k=0,1,  \ldots N$  where  the step-size
$\epsilon$  is  equal to  $2T/N$. Writing $\theta(k)=\theta(t_{(k)})$,
$P(k)=P(t_{(k)})$   and $\bJ(x,k)=\bJ(x,t_{(k)})$, the      transition
amplitude is defined as,
\begin{equation}
T_{fi}[\bJ]\sim\lim_{\epsilon\rightarrow 0}
\int \prod_{k=0}^{N-1}\frac{dP(k)}{(2\pi)}\prod_{k=0}^{N} d\theta(k)
\psi^{*}_{p_{f}}(\theta(N))\psi_{p_{i}}(\theta(0))
\exp \left( i\sum_{k=0}^{N-1} A(k+1,k;\bJ)\right)
\label{discrete}
\end{equation}
where the overall   normalization of   this  expression is  fixed   by
equation (\ref{diagonal}).  The discretized action is given by,
\begin{equation}
A(k+1,k;\bJ)=P(k)\Delta\theta(k)-\epsilon\left[E(P(k))-\int dx\, \bJ(x,k)\cdot
\bvphi(x;P(k),\bar{\theta}(k))\right]
\label{disc}
\end{equation}
where the midpoint and difference fields
$\bar{\theta}(k)$ and $\Delta\theta(k)$, are defined by
\begin{eqnarray}
\bar{\theta}(k)=\frac{1}{2}(\theta(k+1)+\theta(k)) & \qquad{} \qquad{} &
\Delta\theta(k)=\theta(k+1)-\theta(k)
\label{diff}
\end{eqnarray}
respectively.
\paragraph{}
Differentiating the expression  (\ref{discrete}) with  respect to  the
sources $J_{\pm}(x,q_{a})$, where the  grid points labelled by $q_{a}$
are  defined  so  that  $t_{a}=-T+q_{a}\epsilon$, yields a discretized
definition of the Green's functions (\ref{gf}) in which each insertion
of the saddle-point field  configuration is evaluated at the mid-point
value   of  the   collective   coordinate  as  $\varphi_{\pm}(x,q_{a})
=\exp(\pm                          i\bar{\theta}(q_{a}))\phi_{S}\left(
x;\sqrt{m^{2}-\omega(P(q_{a}))^{2}}\right)$.   Each insertion  depends
on the  collective  coordinate  value  only  at two  consecutive  grid
points; $\theta(q_{a})$  and  $\theta(q_{a}+1)$   and  therefore  only
contributes to  the   integral  over    these two  variables.      The
integrations at  remaining grid points  are  unaffected by these terms
and thus, in  the  limit $\epsilon\rightarrow 0$, the   resulting path
integral is divided into segments of the form
\begin{equation}
D(\theta,t;\theta',t')=\int{\cal D}P\int_{\theta(t)=\theta}
^{\theta(t')=\theta'} {\cal D}\theta
\exp\left[i\int_{t}^{t'}dt'' P\dot{\theta}
-H\right]
\label{free}
\end{equation}
describing  free propagation between   initial and  final orientations
$\theta$ and $\theta'$ in the time interval $[t,t']$.  In the simplest
case of the one-point function  between states of definite orientation
the resulting expression involves one insertion of the mid-point field
sandwiched between two such propagators,
\begin{eqnarray}
\langle \theta_{f},+T|\phi_{\pm}(x,t)|\theta_{i},-T \rangle
&\sim &\int d\theta_{-}d\theta_{+}D(\theta_{i},-T;\theta_{-},t)
D(\theta_{+},t;\theta_{f},+T)
\nonumber \\
&&\qquad{} \qquad{}\qquad{}\qquad{}\times\left(\int\frac{dP}{(2\pi)}
\exp(iP\Delta \theta)
\varphi_{\pm}(x;P,\bar{\theta})\right)
\label{onept}
\end{eqnarray}
where             $2\bar{\theta}=(\theta_{+}+\theta_{-})$          and
$\Delta\theta=\theta_{+} -\theta_{-}$.  Using a standard  identity for
operator  products (see    the Appendix of  Ref  \cite{midpoint}),  we
recognize the  field insertion term in the  brackets as an expectation
value  of  the    Weyl  ordered   form     of the field      operator;
$\left\{\hat{\varphi}_{\pm}\right\}_{W}$,
\begin{equation}
\langle \theta_{+}|
\left\{\hat{\varphi}_{\pm}(x;\hat{P},\hat{\theta})\right\}_{W}
|\theta_{-}
\rangle = \int\frac{dP}{(2\pi)}
\exp(iP\Delta \theta)
\varphi_{\pm}(x;P,\bar{\theta})
\label{weyl}
\end{equation}
As usual, the  midpoint definition for   the path integral  yields the
same  results as the Weyl   ordering prescription in the corresponding
operator approach \cite{ber}.
\paragraph{}
Using the standard representation for the
free propagator (\ref{free}) in terms of the
exact eigenfunctions,
\begin{equation}
D(\theta ,t;\theta ',t')=\sum_{p=-\infty}^{+\infty}
\psi^{*}_{p}(\theta ')\psi_{p}(\theta )\exp i(t'-t)E(p)
\label{prop}
\end{equation}
it is straightforward  to evaluate  the  three  residual integrals  in
(\ref{onept}).  After  projecting onto initial   and  final states  of
definite charge, $p_{i}$ and $p_{f}$,  the integrals over $\theta_{+}$
and  $\theta_{-}$ yield $\delta$-functions  which impose $U(1)$ charge
conservation $p_{f}=p_{i}\pm    1$ and also pick   out  a single value
$\bar{p}=(p_{i}+p_{f})/2$  in  the  integration over the  intermediate
angular momentum $P$.  The final result for the one-point function is,
\begin{equation}
\langle p_{f},+T|\phi_{\pm}(x,t)|p_{i},-T\rangle
=\delta_{p_{f},p_{i}\pm 1}n_{i}n_{f}\exp (it\Delta E)\phi_{S}\left(x;
\sqrt{m^{2}-\omega(\bar{p})^{2}}\right)
\label{ans1}
\end{equation}
where $\Delta E=E(p_{i})-E(p_{f})$ and we define the initial and final
state    normalization   factors       $n_{i}(T)=\sqrt{T_{ii}[0]}=\exp
iE(p_{i})T$    and   $n_{f}(T)=\sqrt{T_{ff}[0]}=\exp       iE(p_{f})T$
respectively.   The only non-trivial  feature  of  this result is  the
charge dependent shift in the effective  meson mass which results from
using the correct  saddle-point configuration in the path-integral. As
we will illustrate in  the next section, this  factor is  essential in
order to get a physical contribution to the S-matrix.
\paragraph{}
It  is  straightforward   to   obtain   similar expressions   for  the
saddle-point  contribution to  multi-point Green's functions  directly
from  the discretized  definition of   the  transition amplitude.  For
completeness, we will give the corresponding formula for the $n$-point
function   (\ref{grfs}).   Choosing the time-ordering $-T<t_{1}<\ldots
<t_{n}<+T$, we define  the  intermediate charges and energies  $p_{a}$
and $E_{a}$ at times $t_{a}$ by,
\begin{eqnarray}
p_{a}=p_{i}+\sum_{b=1}^{a}r_{b} & \qquad{} \qquad{} & E_{a}=E(p_{a})
\label{eandp}
\end{eqnarray}
and the corresponding mid-point and difference variables as
\begin{eqnarray}
\bar{p}_{a}=\frac{1}{2}(p_{i}+p_{i-1}) & \qquad{} \qquad{} &
\Delta{E}_{a}=E_{a}-E_{a-1}
\label{deltae}
\end{eqnarray}
The result is,
\begin{equation}
\langle p_{f},+T|\prod_{a=1}^{n}\phi_{s_{a}}(x_{a},t_{a})|p_{i}, -T
\rangle
=\delta_{p_{f},p_{n}}n_{i}n_{f}\prod_{a=1}^{n}
\exp (it_{a}\Delta E_{a})
\phi_{S}\left(x_{a};\sqrt{m^{2}-\omega(\bar{p}_{a})^{2}}\right)
\label{multi3}
\end{equation}
As  for the  one  point function,  the  $\delta$-function  imposes the
conservation   of the total charge, $p_{f}-p_{i}=\sum_{a=1}^{n}r_{a}$,
while the time dependence of the expression is the appropriate one for
energy conservation at each vertex.
\section{The soliton S-matrix}
\paragraph{}
In the previous section we calculated the saddle-point contribution to
Green's functions which   describe the interaction of the   asymptotic
quanta with the   charged  soliton states $|p\rangle$.  The   S-matrix
elements describing any physical  process involving one soliton and an
arbitrary number of mesons can, in principle,  be extracted from these
Green's functions   by the usual LSZ   reduction formula. The simplest
such  process is the  decay  of the charged  soliton state $|p\rangle$
into the state  $|p-1\rangle$ by the  emission of  a single  meson. In
this Section we will calculate the decay  width of the initial soliton
state  to leading  order  in   $1/N$, directly from   the  appropriate
semi-classical one-point function,
\begin{equation}
\langle p-1,+T|\phi_{+}(x,t)|p,-T\rangle
=n_{i}n_{f}\exp (it\Delta E)\phi_{S}\left(x;
\sqrt{m^{2}-\omega(\bar{p})^{2}}\right)
\label{decaygf}
\end{equation}
where $\bar{p}=(p-1/2)$ and $\Delta E=E(p)-E(p-1)$.
\paragraph{}
It is    convenient   to   express  the   soliton    profile  function
$\phi_{S}(x;\mu)$, for  arbitrary mass $\mu$, in  terms  of a momentum
space residue function ${\cal A}(k,\mu)$ according to,
\begin{equation}
\phi_{S}(x;\mu)=\int \frac{dk}{2\pi}\exp(ikx)
\frac{{\cal A}(k;\mu)}{k^{2}+\mu^{2}}
\label{ft}
\end{equation}
where, as we will see below, ${\cal A}(k,\mu)$ is regular and non-zero
at $k^{2}=-\mu^{2}$.  The one-point function for a meson of energy $E$
and   momentum  $k$  is  given   by   the Fourier   transform   of Eqn
(\ref{decaygf}). The Fourier transform with  respect to $t$ yields  an
energy-conserving $\delta$-function while  the transform  with respect
to $x$ follows from the inverse of eqn (\ref{ft});
\begin{equation}
\langle p-1,+T|\phi_{+}(k,E)|p,-T\rangle
=n_{i}n_{f}\delta(E-\Delta E)
\frac{{\cal A}\left(k;\sqrt{m^{2}-\omega(\bar{p})^{2}}\right)}
{k^{2}+m^{2}-\omega(\bar{p})^{2}}
\label{momspace}
\end{equation}
In order for   this one-point function to  contribute  to the S-matrix
element for  the physical  process $|p\rangle  \rightarrow |p-1\rangle
+{\rm  meson}$, it   must  have  a  pole  on   the meson  mass  shell,
$E^{2}=k^{2}+m^{2}$.  Clearly,   the RHS of  (\ref{momspace}) exhibits
such a pole  if   and only if   $\Delta E=\omega(\bar{p})$.   For  the
present purposes we will be content to verify this relation at leading
non-trivial order in $1/N$, in any case loop corrections which we have
so far ignored  will start   to  contribute both  to  $\omega$ and  to
$\Delta E$ at  the  next order.  Expanding  both these   quantities in
powers  of  $1/N$      using (\ref{expansion})   we    find    $\Delta
E=\omega(\bar{p})=(p-1/2)/\Lambda+O(1/N^{2})$ as required.  Hence,  as
advertised in Section 1, the static pole in the soliton background has
been shifted to exactly  the position required by energy conservation.
We can now  apply the standard LSZ  reduction  formula to  the Green's
function   $(\ref{momspace})$.      The     normalization      factors
$n_{i}=\exp(iTE(p))$    and  $n_{f}=\exp(iTE(p-1))$     correspond  to
non-relativistic propagators for the initial  and final soliton states
and it is  appropriate to amputate these  external  legs in  the usual
way. The resulting formula for the momentum space S-matrix element is,
\begin{equation}
{\cal S}_{p}=
\lim_{E^{2}\rightarrow k^{2}+m^{2}}{\cal A}(k;\sqrt{m^{2}-E^{2}})
\label{ss}
\end{equation}
The  behavior of the residue function   ${\cal A}(k;\mu)$ in the limit
$k^{2}\rightarrow -\mu^{2}$ is  determined by  the asymptotic form  of
the  soliton profile at  large  distance. As $|x|\rightarrow\infty$ we
have,  $\phi_{S}(x,\mu)\rightarrow A\exp(-\mu|x|)$    where        the
dimensionless constant $A$ can still depend on $\mu$ through its ratio
with some other   mass  parameter occurring the  original   Lagrangian
(\ref{lag1}).   For convenience,  we will  ignore this model-dependent
complication  and assume that  $A$ is independent  of $\mu$ at leading
order in $1/N$.  By taking the inverse Fourier transform of (\ref{ft})
we  find, ${\cal A}(i\mu,\mu)=2A\mu$.  The S-matrix element (\ref{ss})
therefore has the simple form ${\cal S}_{p}=2Ak$.
\paragraph{}
The  decay width of the  soliton state  $|p\rangle$ is calculated from
the  S-matrix  element   by integrating  over  the  final-state  phase
space. As  we have  neglected  the translational mode of   the soliton
throughout, we  have not  taken  account of the  $O(1/N)$  corrections
arising from the spatial recoil of the soliton.  In this approximation
the differential decay rate is just,
\begin{equation}
d\Gamma_{p}=|{\cal S}_{p}|^{2}\frac{dk}{(2\pi)2E}2\pi\delta(E-\Delta E)
\label{diffrate}
\end{equation}
At leading order in $1/N$ the resulting width is given by,
\begin{equation}
\Gamma_{p}=2A^{2}\sqrt{(p-1/2)^{2}/\Lambda^{2}-m^{2}}
\label{thend}
\end{equation}
This expression depends only on the meson  mass, the soliton moment of
inertia  and the   asymptotic    behavior of  the profile    function,
$\phi_{S}$, and  we stress again  that this result is completely gauge
independent.  For large values of $p$, the width of the excited states
$|p\rangle$  grows linearly  with   the charge simply reflecting   the
increased phase space available for the  decay.  In general, we expect
the state  $|p\rangle$ to appear as a  resonance of width $\Gamma_{p}$
in the  scattering cross-section for a  single meson  off a soliton in
the state  $|p-1\rangle$. Contributions to the  corresponding S-matrix
elements for scattering   processes involving an arbitrary  number  of
mesons  can be obtained by applying  the  LSZ reduction formula to the
general  Green's  function (\ref{multi3}).  In  particular,  the  Born
contribution to the amplitude  for two-body resonant scattering  comes
from  the two-point Green's function $\langle p-1|\phi_{+}\phi_{-}|p-1
\rangle$  in which $\Gamma_{p}$   is self-consistently included as  an
imaginary part for the energy of the intermediate soliton state.
\section{Conclusion}
\paragraph{}
In this paper we have applied the semi-classical method systematically
to   the  interactions  of  mesons with   the  charged  soliton states
associated   with  an  internal   $U(1)$  collective coordinate.    In
particular we have  calculated the complete tree-level contribution to
the one-point  Green's functions describing the   emission of a single
meson and  demonstrated  that  it is  gauge-invariant.   The one-point
function has  a pole on the meson  mass shell which is consistent with
the kinematics of the  physical   decay process.  We  calculated   the
leading semi-classical contribution to the decay widths of the charged
states.  Our  results can  be  interpreted as  the  zeroth-order in  a
semi-classical  loop expansion and   indicate  that it this  expansion
rather than the   standard weak-coupling perturbation theory  which is
appropriate for   the   systematic calculation  of  the  meson-soliton
S-matrix.  However,  in this connection, it  is important to note that
exact cancellations of the  leading-order  contributions can  lead  to
further subtleties  in the   case  of multi-point  Green's  functions.
There are several important omissions in  our discussion of the $U(1)$
model. First,  we have   not  explicitly formulated  the  perturbation
theory in fluctuations around  the saddle-point which is necessary  to
calculate the loop corrections.  Second,  it is necessary to introduce
an explicit collective coordinate to describe the translational motion
of  the  soliton.  As   we have already  discussed,  the corresponding
resummation of all tree graphs   in the translational case   naturally
leads  to     an    appropriately   Lorentz   contracted  saddle-point
configuration. By carefully accounting for the effects of this Lorentz
contraction in  the analysis  given  above, it  should be  possible to
compute the corrections to  the leading-order decay  width due to  the
spatial recoil of the soliton.   Finally, taking into account both the
Born terms and the background scattering terms, all of the above could
be generalized to the two-meson  S-matrix element to give a systematic
semi-classical   expansion of     the   cross-section    for   elastic
meson-soliton scattering in this model.
\paragraph{}
More  generally,  we  expect   any  quantized  theory of  solitons  to
correspond to some    effective    point-like theory  at   least   for
sufficiently   low   energies.  A   well   known  example  of   such a
correspondence  is the mapping  between  the sine-Gordon model and the
Thirring model provided by bosonization \cite{bos}. In this connection
it is interesting to note that  the one-meson S-matrix element, ${\cal
S}_{p}$, calculated in Section 6 has a  simple interpretation in terms
of an effective theory in which the charged soliton states $|p\rangle$
are represented by explicit spinor fields $\psi_{p}$ with Dirac masses
$M(p)=E(p)$.  We  consider the following  effective Lagrangian for the
interactions of the fields $\psi_{p}$ and the charged mesons
\begin{equation}
{\cal L}_{eff}={\cal L}_{\phi}+{\cal L}_{\psi}+{\cal L}_{I}
\label{leff}
\end{equation}
where,
\begin{eqnarray}
{\cal L}_{\phi}&=&\frac{1}{2}(\partial_{\mu} \phi_{+})
(\partial^{\mu}\phi_{+})+\frac{1}{2}(\partial_{\mu} \phi_{-})
(\partial^{\mu}\phi_{-})+\frac{m^{2}}{2}(\phi_{+}^{2}+\phi_{-}^{2}) \\
{\cal L}_{\psi}&=&\sum_{p=-\infty}^{\infty} \bar{\psi}_{p}(i\nd{\partial}
-M(p))\psi_{p} \\
{\cal L}_{I}&=&G\sum_{p=-\infty}^{\infty}\phi_{+}\bar{\psi}_{p}
\gamma_{5}\psi_{p-1}+\phi_{-}\bar{\psi}_{p-1}\gamma_{5}\psi_{p}
\end{eqnarray}
where, in two  space-time dimensions, the  Yukawa coupling $G$ has the
dimensions  of mass.  In  the  limit $N\rightarrow\infty$  the fermion
masses   become    large      and    the      Yukawa       interaction
$G\phi\bar{\psi}\gamma_{5}\psi$   can be   replaced  by  its  standard
non-relativistic                                             reduction
$(G/2M)(\partial_{x}\phi)\bar{\psi}\psi$.      In    this   limit  the
interaction Lagrangian becomes,
\begin{equation}
{\cal L}_{I}\rightarrow
\frac{G}{2M}\sum_{p=-\infty}^{\infty} (\partial_{x}\phi_{+})
\bar{\psi}_{p}\psi_{p-1}+(\partial_{x}\phi_{-})\bar{\psi}_{p-1}\psi_{p}
\label{nonrel}
\end{equation}
The standard Feynman rules derived from this Lagrangian
yield a tree-level amplitude for the decay process $\psi_{p}
\rightarrow \psi_{p-1}+\phi_{+}$ given by $\tilde{{\cal S}}_{p}=(G/2M)k$
where $k$ is the  momentum of the emitted  meson in the center of mass
frame.  Clearly $\tilde{{\cal S}}_{p}={\cal S}_{p}$  for all values of
$p$ if we make  the identification $G=4MA$.  It  is natural to  expect
that this correspondence will remain valid when recoil corrections are
included   and  will also  hold    for the  multi-particle  scattering
amplitudes.
\paragraph{}
The method  presented  here is  directly applicable to  the problem of
$\Delta$-decay in the $SU(2)$ Skyrme model \cite{wip}. We expect that,
as for the $U(1)$ soliton, the net result of systematically accounting
for the  isorotational motion  of the  skyrmion in the  semi-classical
expansion will be to move the static pole  in the Fourier transform of
the skyrmion  background field to exactly  the position required for a
physical  contribution  to the decay amplitude.   Assuming this is the
case, the resulting S-matrix element for $\Delta$-decay would be equal
to the one originally given by Adkins, Nappi  and Witten \cite{ANW} as
a  result of   more general  arguments.    In particular the  proposed
model-independent  relation between  the effective coupling constants,
$g_{\pi  N\Delta}=(2/3)g_{\pi NN}$, would   be obeyed exactly.  In the
$U(1)$ model,  we discovered that  the  widths of the excited  soliton
states increase with the  charge. Assuming a  similar result holds for
the Skyrme model  and the  widths of the  excited baryon  states  grow
rapidly with increasing $SU(2)$ quantum numbers, it is likely that the
predicted $I=J=5/2$  resonance would be too  broad to be distinguished
from  the background  in   pion-nucleon  scattering. This suggests   a
natural  explanation for   the fact  that,  of  the infinite  tower of
large-$N$ baryons, only the two states  of lowest isospin, the nucleon
and   the  $\Delta$, are observed in    nature.  Finally, the possible
correspondence between the soliton S-matrix  elements and an effective
point-like theory is of particular  interest in the Skyrme model where
the latter  would  give an effective  Lagrangian  description  for the
interactions   of baryons  and mesons in    large-$N$ QCD.  We hope to
discuss these interesting  possibilities in  more  detail in the  near
future.
\paragraph{}
The  authors  thank  Leonard  Gamberg   for  helpful  comments  on the
manuscript.   Two of the   authors, ND and   JH, would like to express
their gratitude to the County of Los  Alamos for its hospitality while
part of this work was being completed.

\section*{Appendix A}
\paragraph{}
In this Appendix  we  derive the  choice for $\bpi^{cl}$  given in Eqn
(\ref{cmom})  from the  requirement  that the  change   of phase space
variables     from   $(\bphi,\bpi)$       to       $(\theta,P)$    and
$(\tilde{\bphi},\tilde{\bpi})$         be             a      canonical
transformation. Precisely, we  demand that the  canonical form of  the
Legendre term be preserved  in the  new  variables. This condition  is
expressed in (\ref{legendre}), the LHS of which becomes,
\begin{eqnarray}
\int dx\, \underline{\pi}\cdot\underline{\dot\phi} &= &
\int dx\,\left({\cal M}(\theta)
(\underline{\pi}^{cl}+\underline{\tilde\pi})\right)
\cdot {d\over dt}\left({\cal M}(\theta)\underline{\tilde\phi}\right)
\nonumber \\ &=&
\dot\theta\int dx\,(\underline{\pi}^{cl}+\underline{\tilde\pi})\cdot
\left({\cal M}^T(\theta){\partial {\cal M}(\theta)\over\partial\theta}\right)
\cdot
\underline{\tilde\phi}
+\int dx\,(\underline{\pi}^{cl}+\underline{\tilde\pi})\cdot
\underline{\dot{\tilde\phi}}
\nonumber \\ &=&
\dot\theta\int dx\,(\underline{\pi}^{cl}+\underline{\tilde\pi})\times
\underline{\tilde\phi}\ +\int dx\,
\underline{\tilde\pi}\cdot\underline{\dot{\tilde\phi}}\
+\ \left[{d\over dt}\int dx\,
\underline{\pi}^{cl}\cdot\underline{\tilde\phi}\ -\ \int dx\,
\underline{\dot\pi}^{cl}\cdot\underline{\tilde\phi}\,\right]
\end{eqnarray}
Clearly Eqn (\ref{legendre}) is satisfied if and only if the term in square
brackets in the above equation vanishes and also,
\begin{equation}
P=\int dx\,(\underline{\pi}^{cl}+\underline{\tilde\pi})\times
\underline{\tilde\phi}
\label{iff}
\end{equation}
Remembering the constraint equation (\ref{coneq}), we see that
these criteria are uniquely satisfied by the choice,
\begin{equation}
\bpi^{cl}=\frac{1}{J_{\theta}}\left(P-\int dx\,
\tilde{\bpi}\times\tilde{\bphi}\right)\bff(x)
\label{Acmom}
\end{equation}
It also follows that
\begin{equation}
J_{P}=\Lambda^{-1}\int dx\, \bff\cdot\frac{\delta \bpi^{cl}}{\delta P}
=J_{\theta}^{-1}
\label{Ajacs}
\end{equation}
and therefore  the Jacobian factors   $J_{\theta}$ and $J_{P}$  cancel
exactly in the integrand of (\ref{delta}).
\section*{Appendix B}
\paragraph{}
The purpose of this  Appendix is to find  the simultaneous solution of
the full system  of  saddle-point equations (\ref{phieq}-\ref{coneq}).
The first  step  is  to eliminate all  the   variables other than  the
saddle-point field  itself.   This can   be  accomplished by  a direct
generalization  of   the  corresponding  manipulations   of  equations
(4.20-4.22) in Ref \cite{GJS}.   Taking the scalar product of equation
(\ref{pieq}) with $\bff$ and integrating over space yields,
\begin{equation}
\bar{\nu}=\frac{1}{J_{\theta}}
\left(P-\int dx\,
\tilde{\bpi}\times\tilde{\bphi}\right)
\label{nu}
\end{equation}
We  can   now  eliminate the   Lagrange   multiplier for  the momentum
constraint  from   equation  (\ref{pieq}). Contracting   the resulting
equation  with $\varepsilon_{ik}\tilde{\phi}_{k}$ and integrating with
respect to $x$ we find,
\begin{equation}
\frac{1}{J_{\theta}^{2}}
\left(P-\int dx\,\tilde{\bpi}\times\tilde{\bphi}\right)
=\frac{P}{\Lambda \left(\int dx\, |\tilde{\bphi}|^{2}\right)}
\label{key}
\end{equation}
This relation allows    us to eliminate $\tilde{\bpi}$ from   equation
(\ref{phieq}). Assuming   for the  moment   that  $\bar{\lambda}=0$ in
(\ref{phieq}), the saddle-point equation for $\tilde{\phi}$ becomes,
\begin{equation}
-\frac{\partial^{2} \tilde{\phi}_{S}}{\partial x^{2}} +m^{2}\tilde{\phi}_{S}
+\frac{\delta W}{\delta \phi_{S}} -
\frac{P^{2}\tilde{\phi}_{S}}
{\left( \int dx\, \tilde{\phi}_{S}^{2} \right)^{2}}=0
\label{Amastereq}
\end{equation}
where   $\tilde{\bphi}=(\tilde{\phi}_{S},0)$.    Clearly    the  above
equation coincides with the static classical equation of motion with a
shifted value of the meson mass parameter and is therefore solved by
\begin{equation}
\tilde{\phi}_{S}(x;m)=\phi_{S}
\left(x;\sqrt{m^{2}-P^{2}/\tilde{\Lambda}^{2}}\right)
\label{Ashift}
\end{equation}
where the effective moment of inertia,
$\tilde{\Lambda}=\tilde{\Lambda}(P)$ must satisfy the
self-consistency condition,
\begin{equation}
\tilde{\Lambda}=\int dx\, \left[\phi_{S}\left(x;\sqrt{m^{2}-P^{2}/
\tilde{\Lambda}^{2}}\right)\right]^{2}
\label{Aconsistent}
\end{equation}
Assuming such a value for $\tilde{\Lambda}$ exists, it is easy to show
that            $\tilde{\bphi}^{cl}=(\tilde{\phi}_{S},0)$,         and
$\bar{\lambda}^{cl}=0$ together with the corresponding configurations,
$\tilde{\bpi}^{cl}$  and  $\bar{\nu}^{cl}$,  which are   determined by
equations (\ref{pieq}) and (\ref{nu}), form  a consistent solution  of
the       complete       set     of       saddle-point       equations
(\ref{phieq}-\ref{coneq}). In   particular, because  the  saddle-point
configuration  $\bphi^{cl}$ lives only in the  first  component of the
field vector,  the   first constraint   equation  in (\ref{coneq})  is
trivially satisfied  and  so   the choice   $\bar{\lambda}^{cl}=0$  is
justified {\em a posteriori}.
\paragraph{}
In fact it  can easily be demonstrated  that (\ref{Aconsistent}) has a
real  solution   for  sufficiently  small values   of   the collective
coordinate momentum  $P$.  We define a  generalized  moment of inertia
$\lambda$  as a function of  a variable mass parameter $\mu$ according
to,
\begin{equation}
\lambda(\mu)=\int dx\, [\phi_{S}(x;\mu)]^{2}
\label{lambda}
\end{equation}
{}From the definition, $\lambda(m)=\Lambda$,  and we will further assume
that $\lambda(\mu)$  is  a continuous function  of $\mu$  in the range
$0<\mu\leq m$ with a finite non-zero limit $\lambda(0)=\Lambda_{0}$ as
$\mu\rightarrow   0$.  The    self-consistency  equation  is  given by
$\tilde{\Lambda}=\lambda (\sqrt{m^{2}-P^{2}/\tilde{\Lambda}^{2}})$ and
the  right and  left-hand  sides of   this   equation are  plotted  as
functions   of    $\tilde{\Lambda}$  in   Figure     1.   Noting  that
$\lambda(0)>0$,  we see that for for  $P/m$ less than a critical value
equal   to $\lambda(0)$ the  two   curves  must intersect.  Hence  for
$P/m<\Lambda_{0}$, the  existence of   a     real solution to      the
self-consistency equation  is  guaranteed  by  the continuity  of  the
moment of inertia as the meson mass is varied.
\section*{Figures}
\paragraph{}
Figure 1. A graphical solution of the self-consistency equation
(\ref{consistent}) for the effective moment of inertia $\tilde{\Lambda}$.

\end{document}